\documentclass[twocolumn,prl,showpacs,amsmath,amssymb,superscriptaddress,floatfix]{revtex4}

\usepackage{graphicx}
\usepackage{dcolumn}
\usepackage{bm}
\usepackage{color}

\begin{document}

\title{
Tensor-force-driven Jahn-Teller effect and shape transitions in exotic Si 
isotopes 
}

\author{Yutaka Utsuno}
\affiliation{Advanced Science Research Center, Japan Atomic Energy
Agency, Tokai, Ibaraki 319-1195, Japan}
\author{Takaharu Otsuka}
\affiliation{Department of Physics, University of Tokyo, Hongo,
Bunkyo-ku, Tokyo 113-0033, Japan}
\affiliation{Center for Nuclear Study, University of Tokyo, Hongo,
Bunkyo-ku, Tokyo 113-0033, Japan}
\affiliation{National Superconducting Cyclotron Laboratory, 
Michigan State University, East Lansing MI 48824, USA}
\author{B. Alex Brown}
\affiliation{National Superconducting Cyclotron Laboratory, 
Michigan State University, East Lansing MI 48824, USA}
\affiliation{Department of Physics, Michigan State University, 
East Lansing MI 48824, USA}
\author{Michio Honma}
\affiliation{Center for Mathematical Sciences, University of Aizu, 
Ikki-machi, Aizu-Wakamatsu, Fukushima 965-8580, Japan}
\author{Takahiro Mizusaki}
\affiliation{Institute for Natural Sciences, Senshu University, Tokyo,
  101-8425, Japan}
\author{Noritaka Shimizu}
\affiliation{Center for Nuclear Study, University of Tokyo, Hongo,
Bunkyo-ku, Tokyo 113-0033, Japan}
\date{\today}

\begin{abstract}
We show how the shape evolution of the neutron-rich exotic Si and S
isotopes can be understood as a Jahn-Teller effect that
comes in part from the tensor-driven evolution
of single-particle energies.
The detailed calculations we present are in excellent agreement
with known experimental data, and we point out of new 
features that should be explored in new experiments.
Potential energy surfaces are used to understand the shape evolutions.
The sub-shell closed nucleus, $^{42}$Si, is shown to be a perfect 
example of a strongly oblate shape instead of a sphere through a
robust Jahn-Teller mechanism.   
The distribution of spectroscopic factors measured by $^{48}$Ca(e,e'p)
experiment is shown to be well described, providing a unique test on 
the tensor-driven shell evolution.
\end{abstract}


\maketitle
Among the new frontiers of nuclear physics, one of the most
important is the evolution of single-particle energies
in nuclei far from stability
that led to dramatic changes in the location of magic numbers
and types of collectivity relative to nuclei near stability
\cite{review}.
One of the ingredients is
the tensor-force driven changes in single-particle
energies leading to tensor-force driven shell evolution 
\cite{tensor,vmu}, 
with many experimental examples, {\it e.g.}, \cite{schiffer}.
 
We shall show in this Letter that
tensor-driven shell evolution plays 
a critical role in the rapid shape change
as a function of neutron and/or proton number, including triaxial and 
$\gamma$-unstable shapes. In particular, we show for
the first time \cite{inpc} how this shape change 
at low energy can be 
related to the Jahn-Teller
effect  \cite{jt, reinhard}, where the degenerate intrinsic states
led to a geometric distortion that removes the degeneracy
and lowers the energy.

We study the structure of Si and S isotopes as examples, 
using shell-model calculation with SDPF-MU Hamiltonian 
to be described.  The $sd-pf$ cross-shell 
proton-neutron interaction is important, and is given by
a monopole-based universal interaction
($V_{\rm MU}$) \cite{vmu} except for minor details.
We show a test of this Hamiltonian by the spectroscopy of
K isotopes.  We then demonstrate that
we can describe also the distribution of spectroscopic
factors measured in the (e,e'p) experiment \cite{kramer}, 
exhibiting the first theoretical reproduction of them.
We then show levels, B(E2) values and potential energy surfaces
(PES) of Si and S isotopes for even $N$=22$\sim$28 
in comparison with experiments 
\cite{ensdf, si36, si40, si42, s40, s42, s44}, following 
an intuitive picture of the robust appearance of the oblate
deformation in $^{42}$Si. 
We shall show that not only the property of $^{42}$Si but also 
the rapid change of structure of all these 
Si and S isotopes are naturally described.  

We first outline the details of present shell-model calculations. 
The $sd$ and $pf$ shells are taken as the valence shell, whereas 
no excitation between them is included for the sake of simplicity.
The interactions within each of these shells are based on existing 
interactions: USD \cite{usd} (GXPF1B \cite{gxpf1b}) 
for the $sd$ ($pf$) shell.
Regarding the monopole interaction \cite{vmu,kb3}, 
as pointed out in Ref. \cite{magic}, $V^{T=0}_{0d_{3/2},0d_{5/2}}$ 
must be corrected from its USD value.  We made 
a shift, $\Delta V^{T=0}_{0d_{3/2},0d_{5/2}}
=-3\times\Delta V^{T=1}_{0d_{3/2},0d_{5/2}}=-0.7$ MeV,   
consistently with $V_{\rm MU}$ and SDPF-M \cite{sdpf-m}.  
The monopole- and quadrupole-pairing matrix elements 
$\langle 0f_{7/2} 0f_{7/2} \left| V \right| 0f_{7/2} 0f_{7/2}\rangle_{J=0,2}$ 
are replaced with those of KB3 \cite{kb3} for a better description 
of isotopes of $N \sim$ 20. 
The cross-shell part, most essential for exotic nuclei of 
interest but rather undetermined so far, 
is given basically by $V_{\rm MU}$ of \cite{vmu} with small 
refinements stated below.
The tensor-force component of $V_{\rm MU}$ is nothing but the  
$\pi + \rho$ meson exchange force.  It 
has been used in many cases \cite{tensor, vmu}, and 
has been accounted for microscopically under the new concept of 
Renormalization Persistency \cite{tsunoda}. 
The central-force component is fine tuned from $V_{\rm MU}$ 
with density dependence similar to the one in \cite{sdpota}, 
so that its monopole part becomes closer to that of 
GXPF1 \cite{gxpf1}. 
We include the two-body spin-orbit force of the M3Y 
interaction \cite{m3y}. 
Following USD and GXPF1B, 
all of the two-body matrix elements are scaled by $A^{-0.3}$. 
The single-particle energies (SPE) of the $sd$ shell are taken 
from USD, and those of the $pf$ shell are determined by requesting 
their effective SPEs  
on top of $^{40}$Ca closed shell equal to the single-particle 
energies of GXPF1B. 

The Hamiltonian thus fixed is referred to as SDPF-MU
hereafter, and is diagonalized by the {\sc mshell64} code \cite{mshell}.
Note that none of the cross-shell monopole 
interactions are fitted directly to experiment,  
in contrast to other recent interactions \cite{sdpf-u, kaneko}.

\begin{figure}[t]
 \begin{center}
 \includegraphics[width=6.0cm,clip]{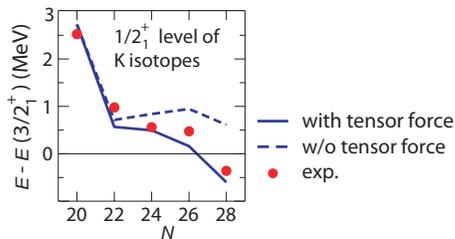}
 \caption{
(Color online) Evolution of $1/2^+_1$ level of K isotopes 
measured from $3/2^+_1$ level. 
Experimental data (filled circles) are compared with theoretical results
including (solid line) or excluding (dashed line) the cross-shell 
tensor force.
}
 \label{fig:k-level}
 \end{center}
\end{figure}

We first study proton single-particle states of odd-$A$ K ($Z$=19) isotopes.
Figure \ref{fig:k-level} exhibits 
$1/2^+_1$ levels for $N$=20$\sim$28.
If valence neutrons occupy the lowest possible orbits 
(filling configuration), 
the $1/2^+_1$ level relative to $3/2^+_1$ corresponds to  
the $0d_{3/2}$-$1s_{1/2}$ gap.
Since this is the case with $N=20$ and 
is almost so with $N=28$, the lowering of the $1/2^+_1$ level 
from $N=20$ to 28 reflects considerable reduction of 
this gap. 
Experimentally observed reduction ($\sim$3 MeV) is reproduced 
remarkably well by SDPF-MU Hamiltonian.
This reduction is due to the proton-neutron monopole
interaction  
$V^{pn}_{0f_{7/2}, 0d_{3/2}}$ that is more attractive than 
$V^{pn}_{0f_{7/2}, 1s_{1/2}}$:   
the difference is 0.44 MeV (at $A=42$), 
out of which the tensor and the central forces contribute, 
respectively, by 0.21 MeV and 0.22 MeV.
The central force yields a stronger attraction between 
$0f_{7/2}$ and $0d_{3/2}$  because of the similarity of their 
radial wave functions \cite{vmu}. 

\begin{figure}[tb]
 \begin{center}
 \includegraphics[width=8.0cm,clip]{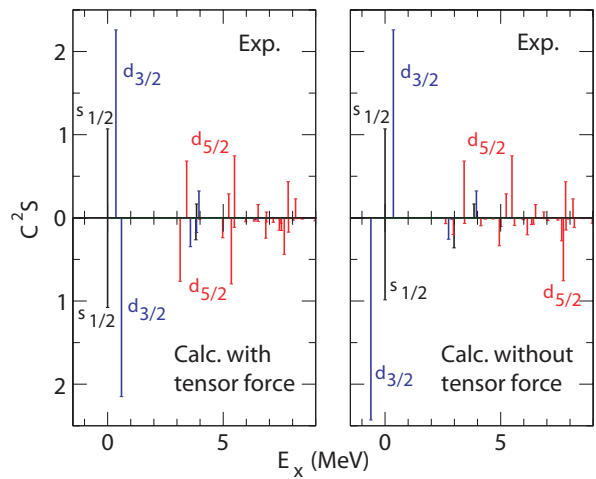}
 \caption{
(Color online) Spectroscopic factors of proton hole states 
measured by $^{48}$Ca(e,e'p) \cite{kramer} (upper) and 
its theoretical calculation (lower left).  The cross-shell tensor force
is removed in lower right panel.     
The black, blue and red bars correspond to $1s_{1/2}$, $0d_{3/2}$ and 
$1d_{5/2}$ states, respectively. 
}
 \label{fig:sfac}
 \end{center}
\end{figure}

We move on to the splitting between the 
$0d_{3/2}$ and $0d_{5/2}$ single-particle energies.
The $0d_{5/2}$ single-particle strength
is highly fragmented due to its large excitation
energy (6-7 MeV).
Spectroscopic factors of 
the proton $sd$-shell orbits have been measured for $^{48}$Ca 
by one-proton removal through ($e, e'p$) reaction \cite{kramer}.
The left panel of Fig.~\ref{fig:sfac} displays 
the experimental values in comparison  
to those obtained by the present calculation, where the usual  
overall quenching factor 0.7 is used \cite{barbieri}. 
The agreement is excellent both in positions of peaks and their 
magnitudes.  
However, this agreement is lost, if the tensor force is removed 
from the cross-shell interaction, as shown in the right panel of 
Fig.~\ref{fig:sfac}.  For instance, the highest $0d_{3/2}$
peak is shifted in the wrong direction, and the main peak of 
$0d_{5/2}$ moves away towards higher energy. 
In the present calculation, as already stated, 
the ESPEs around $^{40}$Ca are consistent with experiments 
with a reasonably large $0d_{3/2}$-$0d_{5/2}$ gap $\sim$7 MeV.  
The proton shell structure evolves from $^{40}$Ca to $^{48}$Ca,
giving rise to the agreement with the fragmentation of 
spectroscopic factors.  
In particular, because 
only the tensor force can change the $0d_{3/2}$-$0d_{5/2}$ gap 
to this order of magnitude (by $\sim$2 MeV), 
the agreement shown in Fig.~\ref{fig:sfac}
provides us with the first evidence from electron scattering
experiments to the tensor-force-driven shell evolution \cite{tensor}.

\begin{figure}[tb]
 \begin{center}
 \includegraphics[width=7.0cm,clip]{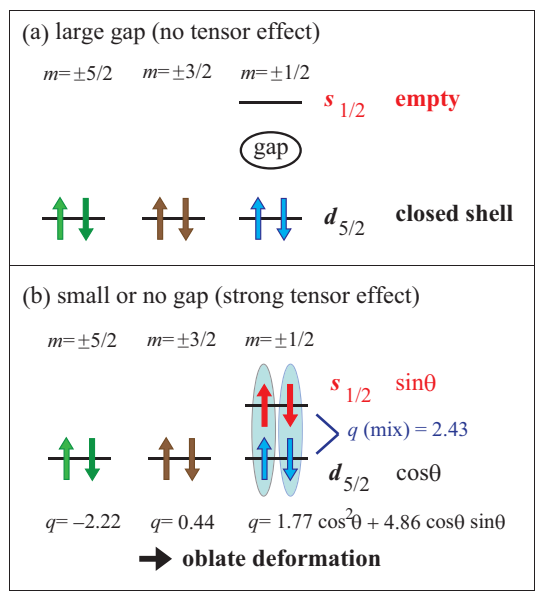}
 \caption{
(Color online) Intuitive illustration of the structure of intrinsic 
state at $Z$=14.  $q$ implies intrinsic quadrupole moment
(fm$^2$).
}
 \label{fig:oblate}
 \end{center}
\end{figure}

We now consider shape transitions driven by 
shell evolution, by taking as an example the exotic Si ($Z$=14) 
isotopes with even $N$=22$\sim$28.
In a conventional view, $Z$=14 is a (sub-)magic number with a large 
$1s_{1/2}$-$0d_{5/2}$ gap, as shown in Fig.~\ref{fig:oblate}(a).
For a large gap, six protons occupy $0d_{5/2}$, forming a closed shell.
This should end up with a spherical shape particularly for a doubly magic 
nucleus, $^{42}$Si ($N$=28),
similar to that observed in $^{34}$Si ($N$=20).
The situation changes when many neutrons occupy the $0f_{7/2}$ orbital.  
Following the mechanism of Otsuka {\it et al.} \cite{tensor}, 
the monopole interaction of the tensor force 
is strongly repulsive between a proton in $0d_{5/2}$ and a neutron 
$0f_{7/2}$.  Hence, as neutrons occupy $0f_{7/2}$, the
spin-orbit splitting decreases and the
energy of proton $0d_{5/2}$ level comes closer to $1s_{1/2}$ level
(The  $1s_{1/2}$ energy is unaffected by this mechanism).

Figure \ref{fig:oblate}(b) presents an intuitive illustration 
on the consequence of this change by taking a simple case 
comprised of $1s_{1/2}$ and $0d_{5/2}$ orbits.
These orbits are mixed, like Nilsson orbits, due to the quadrupole 
deformation of the intrinsic state.
Assuming an axially symmetric deformation, 
single-particle states of the same magnetic quantum 
numbers, denoted $m$, are mixed in the intrinsic states.   
Figure~\ref{fig:oblate} (b) indicates that this occurs for $m=\pm 1/2$ 
between $1s_{1/2}$ and $0d_{5/2}$, 
with amplitudes sin$\theta$ and cos$\theta$, respectively.  
A smaller $1s_{1/2}$-$0d_{5/2}$ gap results in more mixing, while
the phase of mixing amplitude depends on as to whether the shape is 
prolate or oblate.
In the present case, protons occupy the 
states of $m=\pm 5/2, \pm 3/2$, which yield in total a negative 
intrinsic quadrupole moment ({\it i.e.,} an oblate shape).  
The total intrinsic quadrupole moment gains a larger magnitude, 
if the $1s_{1/2}$-$0d_{5/2}$ mixing gives a negative moment.
The proton-neutron interaction, apart from its monopole part, can be 
modeled by a quadrupole-quadrupole interaction, and a similar mixing 
occurs for neutrons in $1p_{3/2}$ and $0f_{7/2}$, producing 
a negative intrinsic moment. 
Thus, by having the mixing leading to a negative intrinsic moment, 
the total magnitude of the moment becomes larger for both protons 
and neutrons, giving rise to stronger binding of the intrinsic
state from the quadrupole-quadrupole interaction.  
The actual structure is determined by the competition between 
actual (effective) gaps and effects of the quadrupole-quadrupole 
interaction, implying that the shapes of Si isotopes depends on how 
the gaps are reduced by the tensor force.  

\begin{figure}[tb]
 \begin{center}
 \includegraphics[width=8.0cm,clip]{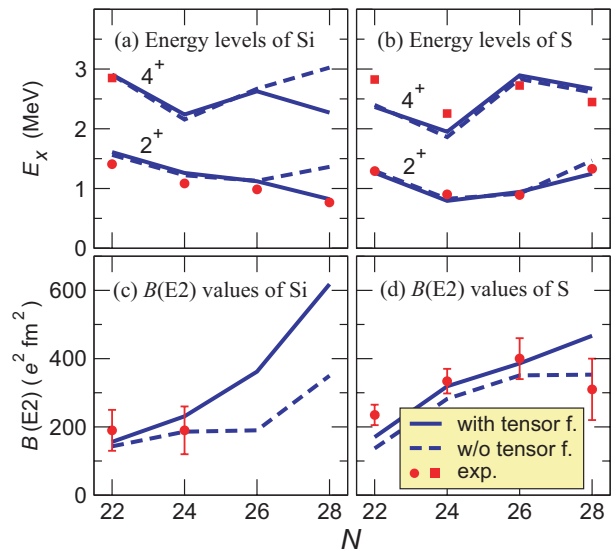}
 \caption{
(Color online) 
(a,b) $2^+_1$ and $4^+_1$ energy levels and (c,d)
$B$(E2;0$^+_1$$\to$2$^+_1$) values of Si and S isotopes 
for $N$=22$\sim$28.
Symbols are experimental data 
\cite{ensdf, si36, si40, si42, s40, s42, s44}.
Solid (dashed) lines are calculations 
with (without) the cross-shell tensor force. 
}
 \label{fig:si-s}
 \end{center}
\end{figure}

We then examine this picture in the context of  
the shell-model calculations with the 
SDPF-MU Hamiltonian.  Figure \ref{fig:si-s} exhibits yrast 
properties of even-$A$ Si and S isotopes.
Effective charges are  $(e_p, e_n)=(1.20e, 0.45e)$  
fixed already by properties of lighter isotopes. 
The overall agreement to experiment is remarkable.
For instance, in the present result, 
$2^+_1$ levels of Si isotopes keep coming down as $N$ increases
consistently with experiment,
whereas some increase is seen at $N=28$ in other shell-model
calculations \cite{sdpf-u, kaneko}. 
The nice agreement suggests that the intuitive picture  
above holds particularly towards $N$=28, resulting in 
strongly deformed shapes with low excitation energies
consistent also with recent measurement by GANIL \cite{si42}.  
In fact, if the tensor force is omitted from the cross-shell 
interaction, $2^+_1$ level of $^{42}$Si$_{28}$ goes up. 
It is of much interest to see the missing experimental data 
in Fig.~\ref{fig:si-s} as well as more precise B(E2) values.
Figure \ref{fig:si-s} exhibits results for S isotopes in good 
overall agreement.

\begin{figure}[t]
 \begin{center}
 \includegraphics[width=7.0cm,clip]{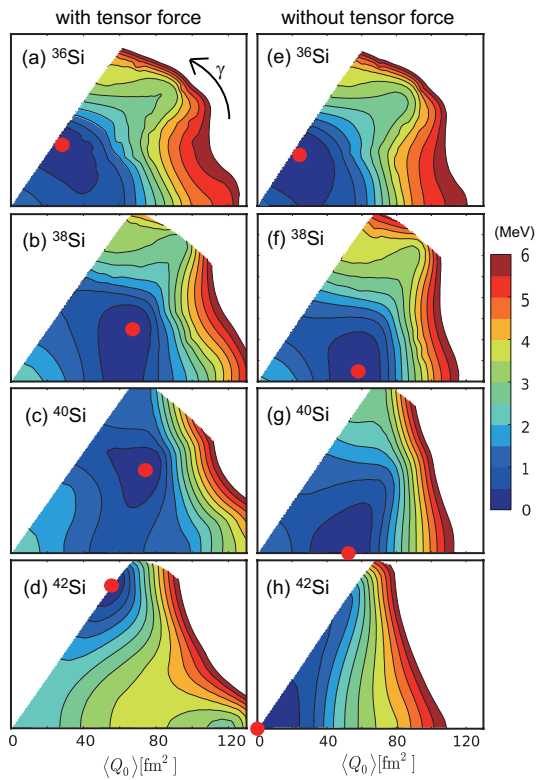}
 \caption{
(Color online) 
Potential energy surfaces of Si isotopes from $N=22$ to 28 
calculated with (left) and without (right) the cross-shell 
tensor force. The energy minima are indicated by red circles.
}
 \label{fig:pes}
 \end{center}
\end{figure}

The potential energy surface (PES) is useful to understand shapes
contained in theoretical calculations.  Figure~\ref{fig:pes} 
exhibits PES for Si isotopes  
obtained by the constrained Hartree-Fock method \cite{ni56} for the 
SDPF-MU Hamiltonian.  The full Hamiltonian is taken in 
Fig.~\ref{fig:pes} (a$\sim$d), whereas the cross-shell tensor force 
is removed in Fig.~\ref{fig:pes} (e$\sim$h).
Shape evolutions are seen very clearly in both sequences 
(a$\sim$d) and (e$\sim$h), starting with similar patterns in $^{36}$Si.
The shape evolves as more neutrons occupy $pf$-shell, with distinct 
differences between the two sequences.
In (b,c), the deformation becomes stronger from (a) with 
triaxial minima, whereas the shape becomes more like modestly prolate
in (f,g).  In (d), one finds a strongly oblate shape with
a sharp minimum, but the minimum is at the spherical shape in (h).
This strong oblate deformation produces low 2$^+$ level and 
large $B$(E2) in Fig.~\ref{fig:si-s} for the ``doubly-closed''
$^{42}$Si. 
Thus, the shape of exotic Si isotopes
changes significantly within the range of 
$\Delta N \sim$6.  This is considered to be a manifestation of 
Jahn-Teller effect with varying shell structure driven by the 
tensor force.
We note that the two sequences produce rather similar levels and
$B$(E2) in Fig.~\ref{fig:si-s} for lighter isotopes, and the 
structure of $^{42}$Si serves a key role in the study of 
the tensor-force effect.  

Although more experiments are needed to clarify present issues, 
there are some hints of the triaxiality from $2^+_2$ level. 
The present calculations with (without) 
the cross-shell tensor force locates it
at 0.62 (1.11) MeV above $2^+_1$ for $^{40}$Si. 
The extraordinary low-lying $2^+_2$ level caused by the tensor force 
appears to agree with a recent $\gamma$-ray experiment \cite{si40}: 
it is proposed that at least either of the observed $\gamma$-rays 
638(8) and 845(6) keV directly feeds the $2^+_1$ state. 

With density-functional methods \cite{Lalazissis,Peru,Rodriguez,Li}, 
Si and S are discussed with somewhat less overall agreement 
with experiment, compared to the present calculation.    
It will be of interest to see single-particle properties 
given by these works in the view of Jahn-Teller effect.
For instance, how much does the $1s_{1/2}$-$0d_{5/2}$ gap change 
from $^{40}$Ca to $^{48}$Ca? 

In summary, we have presented more evidences of the tensor-force 
driven shell evolution, proposed by Otsuka {\it et al.} \cite{tensor}, 
in low-lying states of K isotopes and 
in distribution of spectroscopic factors measured by 
$^{48}$Ca(e,e'p) experiment.  
Similarly, the levels and B(E2)'s of exotic Si and S isotopes are 
described well by the same Hamiltonian.
The nuclear deformation at low excitation energy is a Jahn-Teller
effect, which should be sensitive to the shell structure.
The shell evolution driven by the tensor force plays a crucial
role in rapid shape transitions, including a robust mechanism 
for the appearance of the stable oblate shape at subshell closures
against prolate dominance of the nuclear deformation.  
In future, similarities and differences in $^{78}$Ni will be of 
great interest.  

This work was in part supported by MEXT Grant-in-Aid for 
Scientific Research~(A) 20244022 and for Young Scientists (B) (21740204)
and NSF grant PHY-1068217. 
This work 
is a part of the CNS-RIKEN joint research project on large-scale
nuclear-structure calculations.


\end{document}